\documentclass[letterpaper, 10 pt, conference]{ieeeconf}  
\usepackage[utf8]{inputenc}
\usepackage{amsmath}
\usepackage{cite}
\usepackage{multicol}
\usepackage{graphicx}
\usepackage{array}
\usepackage{multirow}
\usepackage{graphicx}
\usepackage{color}
\usepackage{nomencl}
\usepackage{etoolbox}
\usepackage{amssymb}
\usepackage{comment}
\usepackage{mathtools}
\usepackage{bbold}
\usepackage{yfonts}
\usepackage{soul}
\usepackage{hyperref}
\usepackage[caption=false,font=footnotesize]{subfig}

\IEEEoverridecommandlockouts                              
\newcommand{\sgn}{\texttt{sgn}}

\title{\LARGE \bf
Linear Phase Balancing Scheme using Voltage Unbalance Sensitivities in Multi-phase Power Distribution Grids 
}

\author{Rahul K. Gupta
\thanks{Rahul K. Gupta is with the School of Electrical Engg. \& Computer Science, Washington State University,
        Pullman, WA 99163, USA
        {\tt\small rahul.k.gupta@wsu.edu}}%
}

\begin{document}

\maketitle
\thispagestyle{empty}
\pagestyle{empty}

\begin{abstract}
Power distribution networks, especially in North America, are often unbalanced due to the mix of single-, two-, and three-phase networks as well as due to the high penetration of single-phase devices at the distribution level, such as electric vehicle (EV) chargers and single-phase solar plants. However, the network operator must adhere to the voltage unbalance levels within the limits specified by IEEE, IEC, and NEMA standards for the safety of the equipment as well as the efficiency of the network operation. Existing works have proposed active and reactive power control in the network to minimize imbalances. However, these optimization problems are highly nonlinear and nonconvex due to the inherent nonlinearity of unbalanced metrics and power-flow equations. In this work, we propose a linearization approach of unbalance metrics such as voltage unbalance factors (VUF), phase voltage unbalance rate (PVUR), and line voltage unbalance rate (LVUR) using the first-order Taylor’s approximation. This linearization is then applied to the phase balancing control scheme; it is formulated as a feedback approach where the linearization is updated successively after the active/reactive control setpoint has been actuated and shows improvement in voltage imbalances. We demonstrate the application of the proposed scheme on a standard IEEE benchmark test case, demonstrating its effectiveness.
\end{abstract}
\begin{keywords} 
Phase balancing, voltage unbalance, power distribution networks, linearization.
\end{keywords}
\section{Introduction}
The power distribution systems are progressively facing increased unbalanced power injections due to the rapid growth of single-phase solar plants and residential electric vehicle chargers. This change, along with the presence of inherent unbalancedness (due to a mix of single and two-phase feeders) of North American grids, gives rise to a higher level of voltage unbalance. An increased level of unbalances in distribution systems may trigger protection devices, increase network losses, and damage power equipments (e.g., distribution transformers). A network with high power unbalance results in poor utilization as one of the phases is overloaded compared to others, and this may necessitate system upgrades \cite{ma2020review}. Therefore, the system operators must satisfy the standards regarding the unbalancedness to ensure safe and reliable operation of power distribution networks, defined by IEEE \cite{cooper1987ieee}, IEC \cite{compatibility2009environment}, and NEMA \cite{nema}.


In the existing literature, several methods are proposed to address the issue of voltage imbalance; they can be broadly categorized into three groups. The first method involves investing in new devices, such as static or dynamic VAR compensators, shunt capacitors, tap changers, voltage regulators, etc. \cite{xu2010voltage}, which can be costly and take time to plan and install them. 
The second approach is the phase switching or swapping strategy \cite{zhu1998phase, khodr2006optimal, zhang2024phase}, which involves reassigning certain loads to different phases at a bus to balance loads and reduce voltage unbalance. There are also hybrid approaches that combine phase switching with the use of static VAR compensators \cite{liu2020unbalance}. Utilities often employ these methods manually, where maintenance crews travel to sites and physically switch loads. This process is costly, requiring extensive planning and potentially costing tens of thousands of dollars per phase swap \cite{wang2013phase}. These balancing maneuvers are typically performed periodically, as load characteristics vary with seasonal changes \cite{weckx2014reducing, weckx2015load, geth2015balanced, sun2015phase,girigoudar2023}.
The third method uses active and reactive power control from existing or newly installed solar inverters and other distributed energy resources (DERs). Many studies suggest regulating the active and reactive powers from solar PV inverters~\cite{girigoudar2023,gupta2024improving}. This strategy may require investments in the communication and situational awareness infrastructure for real-time control~\cite{gupta2020grid}. Most of these approaches are usually non-convex due to the non-linear and non-convex nature of the unbalance metrics as well as the power flow equations. Except there are some works that have a proposed successive linearization, such as in \cite{girigoudar2021linearized}, convexification via linearized distflow, e.g., \cite{gupta2025optimizing,taheri2024optimized}, etc.  


In this work, we propose to linearize the voltage unbalance metrics by computing the sensitivities of the voltage unbalance factors with respect to the control variables, enabling to have a linear phase balancing formulation using the first-order Taylor's approximation. These coefficients are derived for different unbalance metrics such as voltage unbalance factors (VUF), phase voltage unbalance rate (PVUR), and line voltage unbalance rate (LVUR). The work shows the accuracy of the linearization on an IEEE benchmark network, then also shows how it can be used to formulate a linearized phase balancing problem. 
In summary, the key contributions of the presented work are (i) analytical derivation of the sensitivities of the different voltage unbalance factors with respect to control variables such as active and reactive powers, (ii) formulating a linearized phase balancing scheme for the active and reactive power control per phase, minimizing the voltage unbalance metrics. 
%
%

The paper is organized as follows. Section~\ref{sec:prelims} introduces the key notations and the voltage magnitude sensitivities. Section~\ref{sec:VUFS} derives the sensitivity coefficients of different voltage unbalance metrics using the voltage magnitude sensitivities. Section~\ref{sec:Application} demonstrates an application of the proposed sensitivity coefficients. Section~\ref{sec:simulation} presents numerical validation, and finally, Section~\ref{sec:conclusion} concludes the work. 
\section{Preliminaries}
\label{sec:prelims}
\subsection{Notations}
We consider a distribution network with network indices $i \in \mathcal{N}$, where the node indices are contained in set $\mathcal{N} = \{1, \dots, N\}$. We consider a three-phase system with phases denoted by $a, b$ and $c$.
The symbols $v_{ia}, v_{ib}$ and $v_{ic}$ denote the complex voltages for phase $ \phi \in \Phi = \{a, b, c\}$ and the symbols $v_{iab}$, $v_{ibc}$ and $v_{ica}$ refer to line voltages. The symbols $p_{ia}, p_{ib}$ and $p_{ic}$ and  $q_{ia}, q_{ib}$ and $q_{ic}$ denote the active and reactive powers, respectively.
The symbols $|y|$ and $y^*$ denote the magnitude and complex conjugate of $y$, respectively. 
\subsection{Voltage Sensitivities}
\label{sec:V_sense}
In the following section, we review the sensitivities of the voltage magnitudes with respect to the control variables, i.e., the active and reactive powers\footnote{The same can be extended to other controllable variables such as transformer taps.}. Let the sensitivity coefficient for complex voltage with respect to a control variable $x$ is denoted by $\mathbf{K}^v_x$, it is
\begin{align}
    \mathbf{K}^v_x = \begin{bmatrix}
        \frac{\partial v_{1a}}{\partial x_{1a}} & \frac{\partial v_{1a}}{\partial x_{1b}}  & \frac{\partial v_{1a}}{\partial x_{1c}}  \dots  \frac{\partial v_{1a}}{\partial x_{Na}} & \frac{\partial v_{1a}}{\partial x_{Nb}}  & \frac{\partial v_{1a}}{\partial x_{Nc}} \\ \\ 
       \frac{\partial v_{1b}}{\partial x_{1a}} & \frac{\partial v_{1b}}{\partial x_{1b}}  & \frac{\partial v_{1b}}{\partial x_{1c}}  \dots  \frac{\partial v_{1b}}{\partial x_{Na}} & \frac{\partial v_{1b}}{\partial x_{Nb}}  & \frac{\partial v_{1b}}{\partial x_{Nc}} \\ \\ 
       \frac{\partial v_{1c}}{\partial x_{1a}} & \frac{\partial v_{1c}}{\partial x_{1b}}  & \frac{\partial v_{1c}}{\partial x_{1c}}  \dots  \frac{\partial v_{1c}}{\partial x_{Na}} & \frac{\partial v_{1c}}{\partial x_{Nb}}  & \frac{\partial v_{1c}}{\partial x_{Nc}} \\ \\ 
        \vdots &  \vdots &  \ddots & \vdots & \vdots \\
        \frac{\partial v_{Na}}{\partial x_{Na}} & \frac{\partial v_{Na}}{\partial x_{1b}}  & \frac{\partial v_{Na}}{\partial x_{1c}}  \dots  \frac{\partial v_{Na}}{\partial x_{Na}} & \frac{\partial v_{Na}}{\partial x_{Nb}}  & \frac{\partial v_{Na}}{\partial x_{Nc}} \\ \\ 
       \frac{\partial v_{Nb}}{\partial x_{1a}} & \frac{\partial v_{Nb}}{\partial x_{1b}}  & \frac{\partial v_{Nb}}{\partial x_{1c}}  \dots  \frac{\partial v_{Nb}}{\partial x_{Na}} & \frac{\partial v_{Nb}}{\partial x_{Nb}}  & \frac{\partial v_{Nb}}{\partial x_{Nc}} \\ \\ 
       \frac{\partial v_{Nc}}{\partial x_{1a}} & \frac{\partial v_{Nc}}{\partial x_{1b}}  & \frac{\partial v_{Nc}}{\partial x_{1c}}  \dots  \frac{\partial v_{Nc}}{\partial x_{Na}} & \frac{\partial v_{Nc}}{\partial x_{Nb}}  & \frac{\partial v_{Nc}}{\partial x_{Nc}}
    \end{bmatrix}
\end{align}
Here, $x$ can be active, reactive power, or other controllable variables such as tap changers. The same coefficient can be defined for voltage magnitudes and angles. 
The sensitivities of voltage magnitudes at bus $i$ and phase $\psi\in{\Phi}$ with respect to control variable at bus $j$, phase $\phi\in\Phi$ can be found as
\begin{align}
    \frac{\partial{|v_{i\psi}|}}{\partial x_{j\phi}} = \frac{1}{|v_{i\psi}|}\Re\Big(v_{i\psi}^*\frac{\partial{v_{i\psi}}}{\partial x_{j\phi}}\Big). 
\end{align}

These sensitivity coefficients can either be computed by inverting the Jacobian of the power-flow solution using the Newton-Raphson approach or by an admittance matrix-based approach, which relies on solving a linear set of equations \cite{christakou2013efficient, maharjan2024generalized, gupta2023quantifying} or an impedance matrix-based approach as described in \cite{zhou2008simplified}.

In the following, we will use the sensitivity of the voltages for deriving the sensitivity of different voltage unbalance metrics. 
\section{Voltage Unbalance Metrics and their Sensitivities}
\label{sec:VUFS}
In a multi-phase electric distribution grid, the voltage unbalance is usually referred to the cases when the three-phase voltages are not symmetrical, which means that either voltage magnitudes are unequal or the voltage phase angles are not 120$^\circ$ apart or both of the above. This may happen due to inherent unbalancedness in the network, such as the presence of single- and two-phase circuits, untransposed circuits, or the presence of single-phase injections (load or generation). The degree of unbalance is quantified by several metrics defined by the international institutions such as IEEE\cite{cooper1987ieee}, IEC \cite{compatibility2009environment} and NEMA \cite{nema}. These standards also prescribe the limits on each of these metrics for safe operation of the power distribution systems.
In practice, the electric distribution system operators try their best to keep the system ``balanced'' within the limits defined by these standards. 

In the following section, we introduce different voltage unbalance metrics and show how their sensitivity with respect to a generic control variable can be analytically expressed.
\subsection{Voltage Unbalance Factor (VUF) and its Sensitivity}
\subsubsection{Definition} 
the VUF is defined as per the IEC standard 61000-2-2 \cite{compatibility2009environment} and is based on the voltage in the sequence domain, which is widely used for unbalance analysis in power systems. Specifically, it is expressed as the ratio of the magnitudes of the negative sequence voltage and positive sequence voltage \cite{pillay2001definitions}. Let denote the VUF$_i$ for node $i$ as $\Xi^\text{VUF}_i$. It is expressed as
\begin{align}
\Xi^\text{VUF}_i = \frac{|v_{i-}|}{|v_{i+}|}
    \label{eq:VUF}
\end{align}
where, 
\begin{align}
\label{eq:sequence}
    \begin{bmatrix}
        v_{i0}\\
        v_{i+}\\
        v_{i-}
    \end{bmatrix}
    =
    \frac{1}{3}\begin{bmatrix}
        1 & 1 & 1 \\
        1 & \alpha & \alpha^2\\
        1 & \alpha^2 & \alpha
    \end{bmatrix}
        \begin{bmatrix}
        v_{ia}\\
        v_{ib}\\
        v_{ic}
    \end{bmatrix}
\end{align}
Here, $\alpha = e^{j\frac{2\pi}{3}}$ and symbols $v_{i0}, v_{i+}$ and $v_{i-}$ denote the zero, positive, and negative sequence voltage, respectively, for $i-$th node. The IEC standard in \cite{compatibility2009environment} mandates keeping the VUF value within 0.02 per unit (or 2\%) for the power distribution systems.

\subsubsection{VUF Sensitivities (VUFS)}
To compute the sensitivity of the VUF of node $i$ with respect to a generic variable $x$ at node $j$ and phase $\phi$, we need to differentiate \eqref{eq:VUF} with respect to $x_{j\phi}$. Therefore, the VUFS can be defined as
\begin{align}
    \text{VUFS}_{ij\phi} = & \frac{\partial(\Xi^\text{VUF}_i)}{\partial(x_{j\phi})} = \\
    = & \frac{1}{|v_{i+}|^2} \Bigg({|v_{i+}|\frac{\partial(|v_{i-}|)}{\partial(x_{j\phi})} - |v_{i-}|\frac{\partial(|v_{i+}|)}{\partial(x_{j\phi})}}\Bigg)
\end{align}
The sensitivities of sequence components $\frac{\partial{v_{i0}}}{\partial x_{j\phi}}, \frac{\partial{v_{i+}}}{\partial x_{j\phi}} , \frac{\partial{v_{i-}}}{\partial x_{j\phi}} $ can be computed by differentiating eq.~\ref{eq:sequence}. 
where the sensitivities of phase voltages $\frac{\partial{v_{ia}}}{\partial x_{j\phi}}, \frac{\partial{v_{ib}}}{\partial x_{j\phi}}, \frac{\partial{v_{ib}}}{\partial x_{j\phi}}$ are the ones defined in Sec.~\ref{sec:V_sense}.
\subsection{Phase Voltage Unbalance Rate (PVUR)}
\subsubsection{Definition}
PVUR is defined by IEEE \cite{cooper1987ieee}, and it uses the phase-to-ground voltages to express the imbalance metric instead of the sequence voltages, as was for the VUF.
It is defined as the ratio of (i) the maximum deviation of the phase voltage magnitudes from their average and (ii) the average phase voltage magnitudes. It is $\Xi_i^\text{PVUR} =$
\begin{align}
     \frac{ \texttt{max}(\big||v_{ia}| -  {v_p}_i^\text{avg}|, ||v_{ib}| -  {v_p}_i^\text{avg}|, ||v_{ic}| -  {v_p}_i^\text{avg}\big|)}{ {v_p}_i^\text{avg}} \label{eq:PVUR}
\end{align}
where, 
$ {v_p}_i^\text{avg} = \frac{1}{3}(|v_{ia}| + |v_{ib}| + |v_{ic}|).$

As per the IEEE standard 141-1993 \cite{cooper1987ieee}, the PVUR should be below 0.02 per unit or 2 \%.

The PVUR expression can be simplified into three cases as, 
PVUR  = 
\begin{align}
\label{eq:PVUR2}
\begin{dcases}
    \frac{\big|2|v_{ia}| - |v_{ib}| - |v_{ic}|\big|}{|v_{ia}| + |v_{ib}|+ |v_{ic}|} \\ \qquad  \quad \big||v_{ia}| - {v_p}_i^\text{avg} \big| > \big||v_{ib}| - {v_p}_i^\text{avg} \big| ~\&~ \big||v_{ic}| - {v_p}_i^\text{avg} \big| \\
    \frac{\big|2|v_{ib}| - |v_{ia}| - |v_{ic}|\big|}{|v_{ia}| + |v_{ib}|+ |v_{ic}|}  \\ \qquad  \quad \big||v_{ib}| - {v_p}_i^\text{avg} \big| > \big||v_{ia}| - {v_p}_i^\text{avg} \big| ~\&~\big||v_{ic}| - {v_p}_i^\text{avg} \big|\\
    \frac{\big|2|v_{ic}| - |v_{ia}| - |v_{ic}|\big|}{|v_{ia}| + |v_{ib}|+ |v_{ic}|}  \\ 
    \quad  \qquad  \big||v_{ic}| - {v_p}_i^\text{avg} \big| > \big||v_{ia}| - {v_p}_i^\text{avg} \big| ~\&~ \big||v_{ib}| - {v_p}_i^\text{avg} \big|.\\
\end{dcases}
\end{align}
\subsubsection{PVUR Sensitivities (PVURS)}
\label{eq:PVURS_eq}
Based on the three cases defined in \eqref{eq:PVUR2}, the PVUR sensitivity for each of these cases can be computed by differentiating with respect to the control variables. Here, $\phi, \psi$ varies over the phase set $\Phi$ defined earlier.

The PVURS for the \textit{first case} ($\big||v_{ia}| - {v_p}_i^\text{avg}\big| > \big||v_{ib}| - {v_p}_i^\text{avg}\big| ~\&  ~\big||v_{ic}| - {v_p}_i^\text{avg}\big|$)) can be expressed as, $\text{PVUR}_{ij\phi}=$
\begin{align}
\begin{aligned}
& \frac{\partial(\Xi_i^\text{PVUR} )}{\partial x_{j\phi}} = 
     \frac{1}{\big(\sum_{\psi} {|v_{i\psi}|}\big)^2}\Big\{\big(\sum_{\psi} {|v_{i\psi}|}\big) \times \sgn (2|v_{ia}| - \\
     & |v_{ib}| - |v_{ic}|) \times   \Big(2\frac{\partial |v_{ia}|}{\partial x_{j\phi}} - \frac{\partial |v_{ib}|}{\partial x_{j\phi}} -  \frac{\partial |v_{ic}|}{\partial x_{j\phi}}\Big)  - \\
     & \big|2|v_{ia}| - |v_{ib}| - |v_{ic}|\big| \times \Big(\sum_{\psi} \frac{\partial {|v_{i\psi}|}}{ \partial x_{j\phi}}\Big)\Big\}
\end{aligned}
\end{align}
where, $\sgn$ refers to the signum function.

Similarly for the \textit{second case} ($\big||v_{ib}| - {v_p}_i^\text{avg}\big| > \big||v_{ia}| - {v_p}_i^\text{avg}\big| ~\&~\big||v_{ic}| - {v_p}_i^\text{avg}\big|$), the $\text{PVUR}_{ij\phi}=$
\begin{align}
\begin{aligned}
& \frac{\partial(\Xi^\text{PVUR} )}{\partial x_{j\phi}} = 
     \frac{1}{\big(\sum_{\psi} {|v_{i\psi}|}\big)^2}\Big\{\big(\sum_{\psi} {|v_{i\psi}|}\big) \times\sgn(2|v_{ib}| - \\
     &  |v_{ia}| - |v_{ic}|) \times   \Big(2\frac{\partial |v_{ib}|}{\partial x_{j\phi}} - \frac{\partial |v_{ia}|}{\partial x_{j\phi}} -  \frac{\partial |v_{ic}|}{\partial x_{j\phi}}\Big)  - \\
     & \big|2|v_{ib}| - |v_{ia}| - |v_{ic}|\big| \times \Big(\sum_{\psi} \frac{\partial {|v_{i\psi}|}}{ \partial x_{j\phi}}\Big)\Big\},
\end{aligned}
\end{align}
and \textit{third case} $\big||v_{ic}| - {v_p}_i^\text{avg}\big| > \big||v_{ia}| - {v_p}_i^\text{avg}\big| ~\&~ \big||v_{ib}| - {v_p}_i^\text{avg}\big|$, it is defined as, $\text{PVUR}_{ij\phi}=$
\begin{align}
\begin{aligned}
& \frac{\partial(\Xi^\text{PVUR} )}{\partial x_{j\phi}} = 
     \frac{1}{\big(\sum_{\psi} {|v_{i\psi}|}\big)^2}\Big\{\big(\sum_{\psi} {|v_{i\psi}|}\big) \times \\
     & \sgn{(2|v_{ic}| - |v_{ia}| - |v_{ib}|)} \times   \Big(2\frac{\partial |v_{ic}|}{\partial x_{j\phi}} - \frac{\partial |v_{ia}|}{\partial x_{j\phi}} -  \frac{\partial |v_{ib}|}{\partial x_{j\phi}}\Big)  - \\
     & \big|2|v_{ic}| - |v_{ia}| - |v_{ib}|\big| \times \Big(\sum_{\psi} \frac{\partial {|v_{i\psi}|}}{ \partial x_{j\phi}}\Big)\Big\}.
\end{aligned}
\end{align}

\subsection{Line Voltage Unbalance Factor (LVUR)}
\subsubsection{Definition}
It is defined by NEMA \cite{nema} and used mostly by the drives and motor manufacturers. LVUR is defined in a similar way as PVUR, but with the line voltages. The LVUR is defined as
\begin{align}
    \Xi_i^\text{LVUR} = \frac{\texttt{max}(||v_{iab}| - {v_l}_i^\text{avg}|, ||v_{ibc}| - {v_l}_i^\text{avg}|, ||v_{ica}| - {v_l}_i^\text{avg}|)}{{v_l}_i^\text{avg}}
     \label{eq:LVUR}
\end{align}
where, ${v_l}_i^\text{avg} = \frac{1}{3}(|v_{iab}| + |v_{ibc}| + |v_{ica}|)$.
LVUR can be rewritten into three cases as
LVUR  = 
\begin{align}
\label{eq:LVUR2}
\begin{dcases}
    \frac{\big|2|v_{iab}| - |v_{ibc}| - |v_{ica}|\big|}{|v_{iab}| + |v_{ibc}|+ |v_{ica}|} \\ \qquad \quad \big||v_{iab}| - {v_l}_i^\text{avg}\big| > \big||v_{ibc}| - {v_l}_i^\text{avg}\big| ~\&~ \big||v_{ica}| - {v_l}_i^\text{avg}\big|\\
    \frac{\big|2|v_{ibc}| - |v_{iab}| - |v_{ica}|\big|}{|v_{iab}| + |v_{ibc}|+ |v_{ica}|} \\ \qquad \quad \big||v_{ibc}| - {v_l}_i^\text{avg}\big| > \big||v_{iab}| - {v_l}_i^\text{avg}\big| ~\&~ \big||v_{ica}| - {v_l}_i^\text{avg}\big|\\
    \frac{\big|2|v_{ica}| - |v_{iab}| - |v_{ibc}|\big|}{|v_{iab}| + |v_{ibc}|+ |v_{ica}|} \\ \qquad \quad \big||v_{ica}| - {v_l}_i^\text{avg}\big| > \big||v_{iab}| - {v_l}_i^\text{avg}\big| ~\&~ \big||v_{ibc}| - {v_l}_i^\text{avg}\big|.\\
\end{dcases}
\end{align}

\subsubsection{LVUR sensitivities (LVURS)} 
\label{eq:LVURS_eq}
can be computed similarly to PVURS. The key difference here is that we need the sensitivity of line voltage magnitudes. The sensitivity of line voltages for phase $\phi, \psi, \psi' \in \Phi$ as $v_{i\psi\psi'}$ (complex) can be computed as 
    $\frac{\partial v_{i\psi\psi'}}{\partial x_{j\phi}} = \frac{\partial v_{i\psi}}{\partial x_{j\phi}}  - \frac{\partial v_{i\psi'}}{\partial x_{j\phi}} $
and the sensitivity of line voltage magnitudes can be computed as
\begin{align}
    \frac{\partial{|v_{i\psi\psi'}|}}{\partial x_{j\phi}} = \frac{1}{|v_{i\psi\psi'}|}\Re\Big(v_{i\psi\psi'}^*\frac{\partial{v_{i\psi\psi'}}}{\partial x_{j\phi}}\Big).
\end{align}

Then, the LVURS can be derived for \textit{first case} $(\big||v_{iab}| - {v_l}_i^\text{avg}\big| > \big||v_{ibc}| - {v_l}_i^\text{avg}\big| ~\&~ \big||v_{ica} - {v_l}_i^\text{avg}|)$ in \eqref{eq:LVUR2} as, LVURS$_{ij\phi} = $
\begin{small}
\begin{align}
\begin{aligned}
& \frac{\partial(\Xi_i^\text{LVUR})}{\partial x_{j\phi}} = 
     \frac{1}{\big(\sum_{\psi,\psi', \psi \neq \psi'} {|v_{i\psi\psi'}|}\big)^2}\Bigg\{\Big(\sum_{\psi,\psi', \psi \neq \psi'} {|v_{i\psi\psi'}|}\Big) \times \\ &  \sgn{(2|v_{iab}| - |v_{ibc}| - |v_{ica}|)} \times  \Bigg(2\frac{\partial |v_{iab}|}{\partial x_{j\phi}} - \frac{\partial |v_{ibc}|}{\partial x_{j\phi}} -  \\ &\frac{\partial |v_{ica}|}{\partial x_{j\phi}}\Bigg)   - \big|2|v_{iab}| - |v_{ibc}| - |v_{ica}|\big| 
     \times \Bigg(\sum_{\psi,\psi', \psi \neq \psi'} \frac{\partial {|v_{i\psi\psi'}|}}{ \partial x_{j\phi}}\Bigg)\Bigg\}.
\end{aligned}
\end{align}
\end{small}
The LVURs for \textit{second case} $)\big||v_{ibc}| - {v_l}_i^\text{avg}\big| > \big||v_{iab}| - {v_l}_i^\text{avg}\big| ~\&~ \big||v_{ica}| - {v_l}_i^\text{avg}\big|)$, can be expressed as, LVURS$_{ij\phi} = $

\begin{small}
\begin{align}
\begin{aligned}
& \frac{\partial(\Xi_i^\text{LVUR} )}{\partial x_{j\phi}} = 
     \frac{1}{\big(\sum_{\psi,\psi', \psi \neq \psi'} {|v_{i\psi\psi'}|}\big)^2}\Bigg\{\Big(\sum_{\psi,\psi', \psi \neq \psi'} {|v_{i\psi\psi'}|}\Big) \times \\ &  \sgn{(2|v_{ibc}| - |v_{iab}| - |v_{ica}|)} \times  \Bigg(2\frac{\partial |v_{ibc}|}{\partial x_{j\phi}} - \frac{\partial |v_{iab}|}{\partial x_{j\phi}} -  \\ & \frac{\partial |v_{ica}|}{\partial x_{j\phi}}\Bigg)  - \big|2|v_{ibc}| - |v_{iab}| - |v_{ica}|\big| 
     \times \Bigg(\sum_{\psi,\psi', \psi \neq \psi'} \frac{\partial {|v_{i\psi\psi'}|}}{\partial x_{j\phi}}\Bigg)\Bigg\}.
\end{aligned}
\end{align}
\end{small}

The LVURs for the for the \textit{third case} $(\big||v_{ica}| - {v_l}_i^\text{avg}\big| > \big||v_{iab}| - {v_l}_i^\text{avg}\big| ~\&~ \big||v_{ibc}| - {v_l}_i^\text{avg}\big|)$, can be expressed as, LVURS$_{ij\phi} = $

\begin{small}
\begin{align}
\begin{aligned}
& \frac{\partial(\Xi_i^\text{LVUR} )}{\partial x_{j\phi}} = 
     \frac{1}{\big(\sum_{\psi,\psi', \psi \neq \psi'} {|v_{i\psi\psi'}|}\big)^2}\Bigg\{\Big(\sum_{\psi,\psi', \psi \neq \psi'} {|v_{i\psi\psi'}|}\Big) \times \\ &  \sgn{(2|v_{ica}| - |v_{ibc}| - |v_{iab}|)} \times  \Bigg(2\frac{\partial |v_{ica}|}{\partial x_{j\phi}} - \frac{\partial |v_{ibc}|}{\partial x_{j\phi}} -  \\ & \frac{\partial |v_{iab}|}{\partial x_{j\phi}}\Bigg)  - \big|2|v_{ica}| - |v_{ibc}| - |v_{iab}|\big| 
     \times \Bigg(\sum_{\psi,\psi', \psi \neq \psi'} \frac{\partial {|v_{i\psi\psi'}|}}{\partial x_{j\phi}}\Bigg)\Bigg\}.
\end{aligned}
\end{align}
\end{small}

In the following, the sensitivity of the unbalance metrics is used for voltage balancing.
\section{Linear Voltage Balancing Problem}
\label{sec:Application}

\subsection{Linearization using sensitivity coefficients}
In this section, we show the use of voltage unbalance sensitivities for linearized expression of voltage unbalance factors. Let denote the generic voltage unbalance metrics by $(\Xi_i^{\theta})$ where $\theta = \{\text{VUF}, \text{PVUR}, \text{LVUR}\}$; using the sensitivities defined in the previous section, they can be expressed as a linear expression for each node $i \in \mathcal{N}$ using the first order Taylor's approximation. Assuming that the control variables, i.e., the power injection are denoted by $\mathbf{x}$, then the voltage unbalance metrics can be computed as 
\begin{align}
    \Xi^{\theta}_i(\mathbf {x + dx}) \approx \Xi^{\theta}_i( {\mathbf x}) + \sum_j\sum_{\phi}\frac{\partial( \Xi^{\theta}_i(\mathbf x ))}{\partial x_{j\phi}} (dx_{j\phi})
    \label{eq:linearization}
\end{align}
where $dx_{j\phi}$ denote the active/reactive power deviation at node $j$, phase $\phi$ and ${\mathbf{x}}$ denote the current operating point.

In a similar way, the voltage magnitudes can be linearized using the voltage sensitivity coefficients \cite{gupta2022model} as
\begin{align}
    |v_{i\psi}(\mathbf {x+dx})| & \approx |v_{i\psi}({\mathbf x})| + \sum_j\sum_{\phi} \frac{\partial |v_{i\psi}(\mathbf x )| }{\partial x_{j\phi}}(dx_{j\phi}).
    \label{eq:Vlinearization}
\end{align}

In the following section, the linearized equations for the unbalance metrics \eqref{eq:linearization} and voltage magnitudes \eqref{eq:Vlinearization} will be used to formulate the phase balancing problem.
\subsection{Phase balancing optimization problem}
\label{sec:phase_balancing_application}
In a three-phase unbalanced system, phase balancing mechanism usually refers to the balancing of the voltage magnitudes, i.e., the three-phases become nearly balanced such that the overall unbalancedness complies with the limits defined by IEEE \cite{cooper1987ieee}, IEC \cite{compatibility2009environment} or NEMA \cite{nema} as defined in Sec.~\ref{sec:VUFS}. 
In this specific case, we demonstrate the phase balancing application using active/reactive power control. Let consider that at each phase ($\phi$) of each node ($j$), there is flexibility to modify the injection by $dx_{j\phi}$ which is bounded by a factor $\beta \in [0, 0.05]$ of the nominal base load per phase per node which means that the active or reactive power flexibility is limited by 5\%.

Then, the phase balancing problem can be formulated as an optimization problem with an objective to minimize the voltage unbalance metric and the grid constraints. It can be expressed as
\begin{subequations}
\label{eq:phase_balance_op}
\begin{align}
    \texttt{minimize}  & ~\sum_{i\in\mathcal{N}}\Xi^{\theta}_i(\mathbf {x+dx}) \\
   \texttt{subject to:} ~ & 0 \leq \Xi^{\theta}_i(\mathbf {x+dx})\leq \overline{\Xi}^{\theta} ~~~ \forall i \in \mathcal{N}\\
   & \underline{v} \leq  |v_{i\phi}(\mathbf {x+dx})| \leq \overline{v} ~~~\forall i \in \mathcal{N}\\
   & -\beta \hat{\mathbf{x}} \leq \mathbf{dx} \leq \beta \hat{\mathbf{x}} \label{eq:x_bounds}\\
    & \sum_{\phi} dx_{j\phi} =0 ~~~ \forall j \in \mathcal{N} \label{eq:power_cons}.
\end{align}
\end{subequations}
where $\overline{\Xi}^{\theta}$ is the upper unbalance metric defined by standards \cite{cooper1987ieee, compatibility2009environment, nema} which in most cases is 0.02 per unit as also described in Sec .~\ref {sec:VUFS}. The symbols $\underline{v}$ and $ \overline{v}$ are lower and upper bounds on the voltage magnitudes, which are set as 0.95 and 1.05 per unit, respectively. 
Eq.~\eqref{eq:x_bounds} are the limits on the controllable power flexibility per phase per node, where $\hat{\mathbf{x}}$ refers to the nominal injection. 
Eq.~\eqref{eq:power_cons} ensures that the sum of the new power injections matches the net demand per node in the original system, which means that the net sum of flexible deviations is zero.

As it can be seen, this phase balancing formulation is a linear program (LP), thanks to the linearized expression in \eqref{eq:linearization} and \eqref{eq:Vlinearization}. The proposed formulation will be validated in the following section on an IEEE benchmark network.

\section{Numerical Validation}
\label{sec:simulation}
The proposed scheme on phase balancing is validated on IEEE 34-bus system, the network parameters and the power injections are detailed in \cite{kersting2010comprehensive}.
\begin{figure}[!htbp]
    \centering
    \vspace{-0.5em}
    \includegraphics[width=0.95\linewidth]{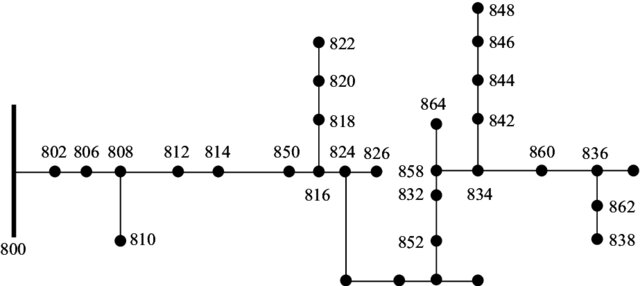}
    \caption{IEEE 34 node test feeder (\emph{simplified)}.}
    \label{fig:IEEE34}
\end{figure}
\begin{figure*}[!htbp]
    \centering
    \subfloat[$\partial (\Xi^\text{VUF})/ \partial p_a$]{ \includegraphics[width=0.16\linewidth]{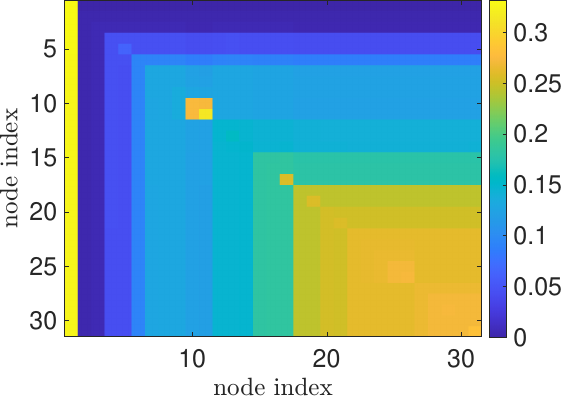}} 
    \subfloat[$\partial (\Xi^\text{VUF})/ \partial p_b$]{ \includegraphics[width=0.16\linewidth]{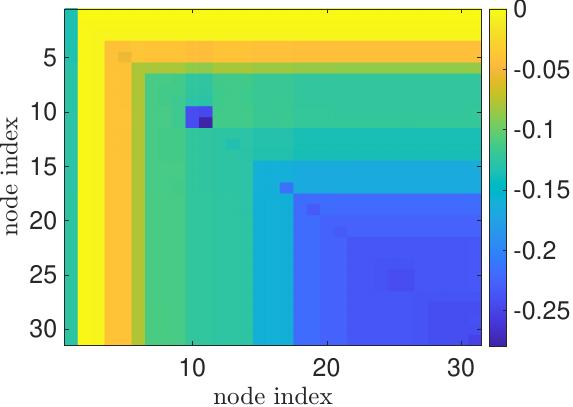}} 
    \subfloat[$\partial (\Xi^\text{VUF})/ \partial p_c$]{ \includegraphics[width=0.16\linewidth]{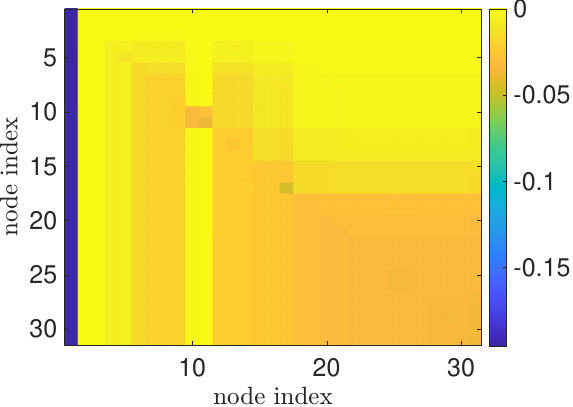}} 
    \subfloat[$\partial (\Xi^\text{VUF})/ \partial q_a$]{ \includegraphics[width=0.16\linewidth]{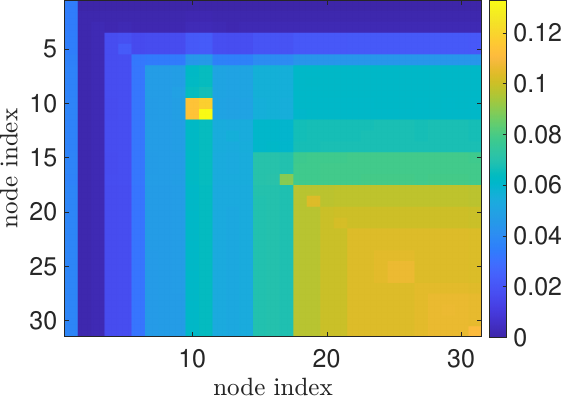}} 
    \subfloat[$\partial (\Xi^\text{VUF})/ \partial q_b$]{ \includegraphics[width=0.16\linewidth]{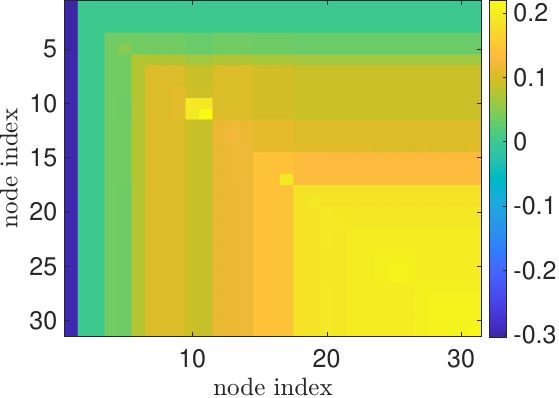}} 
    \subfloat[$\partial (\Xi^\text{VUF})/ \partial q_c$]{ \includegraphics[width=0.16\linewidth]{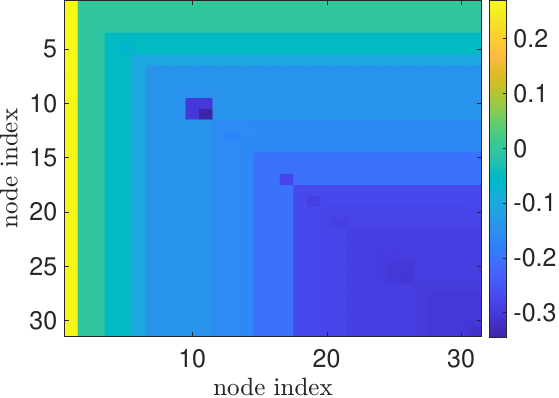}} \\
    \subfloat[$\partial (\Xi^\text{PVUR})/ \partial p_a$]{ \includegraphics[width=0.16\linewidth]{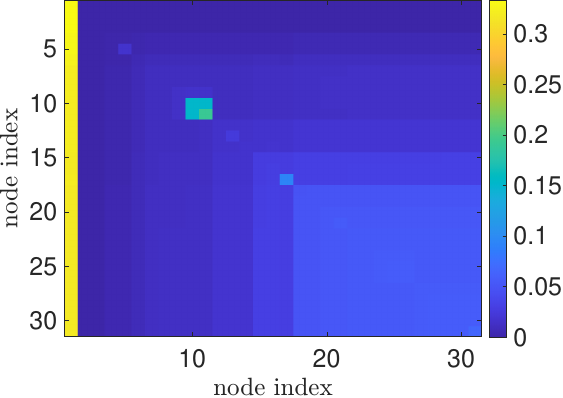}} 
    \subfloat[$\partial (\Xi^\text{PVUR})/ \partial p_b$]{ \includegraphics[width=0.16\linewidth]{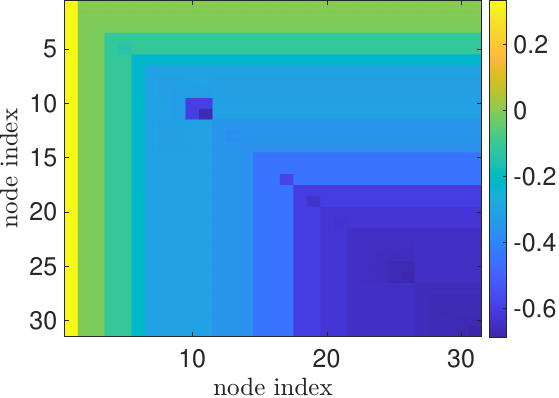}}
    \subfloat[$\partial (\Xi^\text{PVUR})/ \partial p_c$]{ \includegraphics[width=0.16\linewidth]{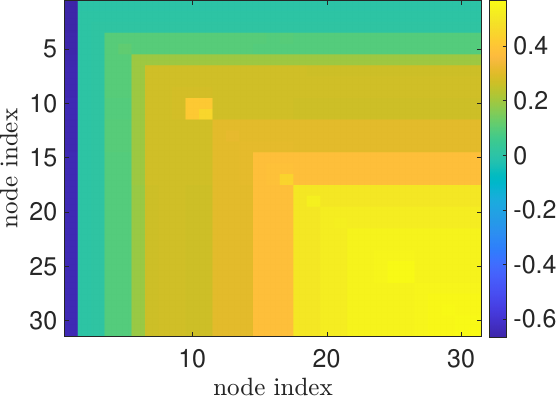}}
    \subfloat[$\partial(\Xi^\text{PVUR})/ \partial q_a$]{ \includegraphics[width=0.16\linewidth]{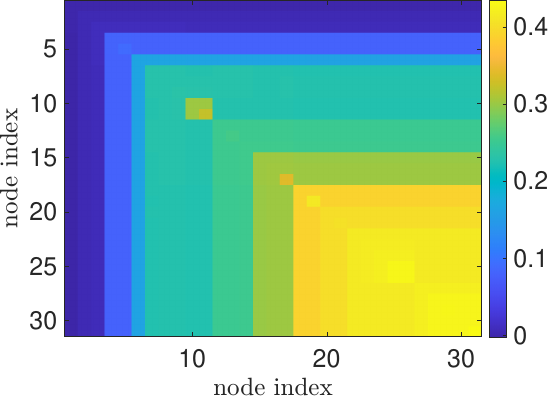}}
    \subfloat[$\partial(\Xi^\text{PVUR})/ \partial q_b$]{ \includegraphics[width=0.16\linewidth]{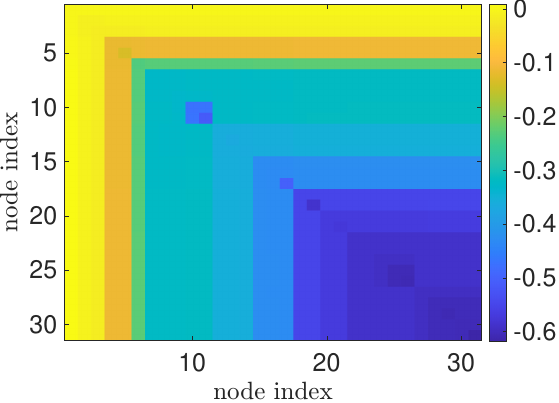}} 
    \subfloat[$\partial(\Xi^\text{PVUR})/ \partial q_c$]{ \includegraphics[width=0.16\linewidth]{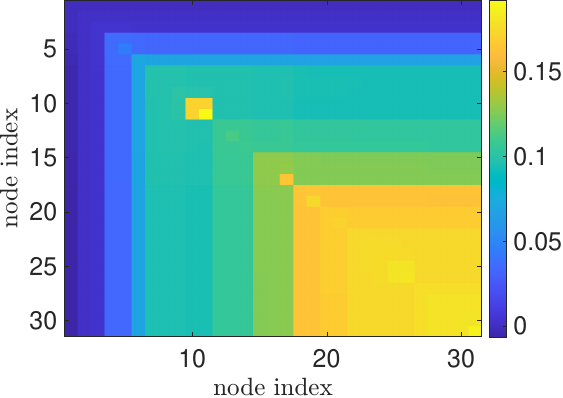}} \\
    \subfloat[$\partial (\Xi^\text{LVUR})/ \partial p_a$]{ \includegraphics[width=0.16\linewidth]{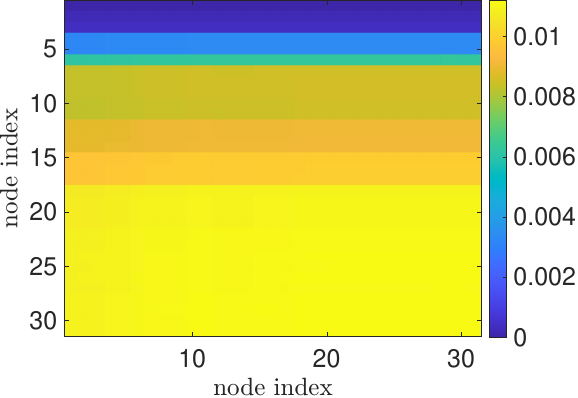}} 
    \subfloat[$\partial (\Xi^\text{LVUR})/ \partial p_b$]{ \includegraphics[width=0.16\linewidth]{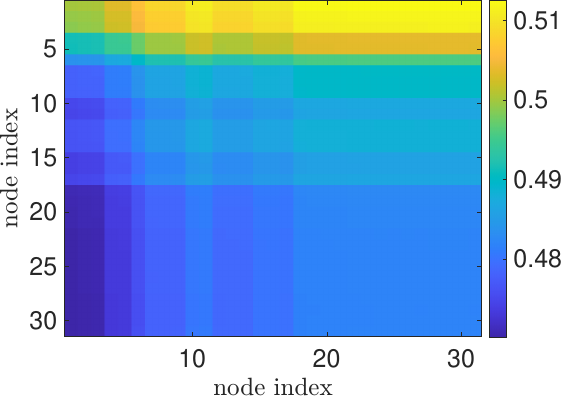}} 
    \subfloat[$\partial (\Xi^\text{LVUR})/ \partial p_c$]{ \includegraphics[width=0.16\linewidth]{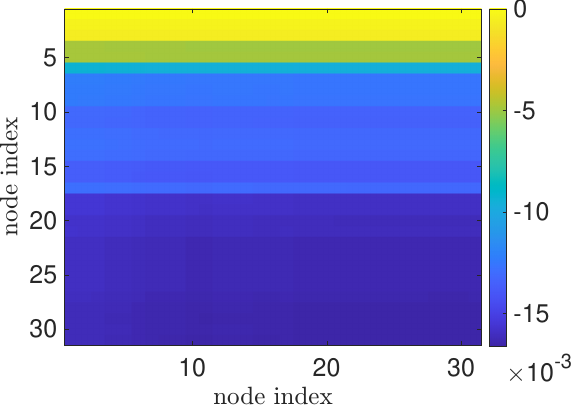}}
    \subfloat[$\partial(\Xi^\text{LVUR})/ \partial q_a$]{ \includegraphics[width=0.16\linewidth]{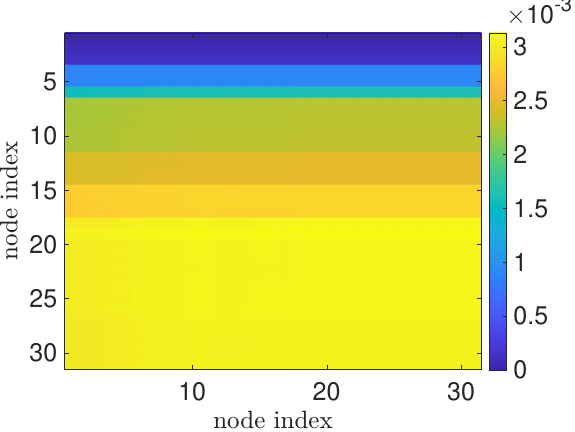}} 
    \subfloat[$\partial(\Xi^\text{LVUR})/ \partial q_b$]{ \includegraphics[width=0.16\linewidth]{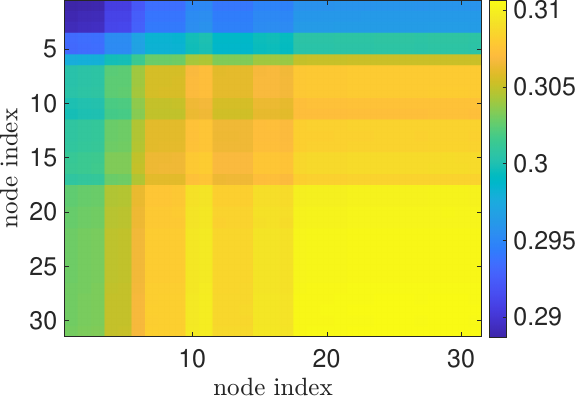}}
    \subfloat[$\partial(\Xi^\text{LVUR})/ \partial q_c$]{ \includegraphics[width=0.16\linewidth]{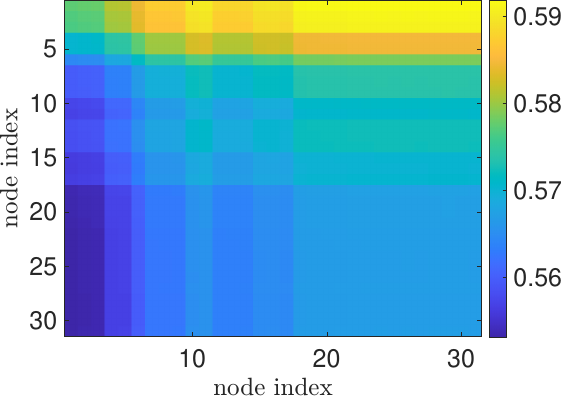}}
    \caption{Sensitivities: VUFS, PVURS and LVURS with respect to active and reactive power injections at phases $a, b$ and $c$.}
    \label{fig:VUFSPQ}
\end{figure*}

In the following, first, we present the results for sensitivity coefficient assuming a nominal demand (Sec.~\ref{sec:sense_w_nominal}), then we vary the injections for a particular phase and node to observe its effect on the sensitivity coefficient (Sec.~\ref{sec:sense_sense}), then finally, we present the result of phase balancing scheme (Sec.~\ref{sec:phase_bal_results}).
\subsection{Sensitivity coefficients per phase per node}
\label{sec:sense_w_nominal}
Figure~\ref{fig:VUFSPQ} shows VUFS, PVURS, and LVURS for the IEEE-34 case in the form of a heatmap plot. These sensitivities are computed for each node with respect to the active and reactive power injections of different nodes and phases in Figs~\ref{fig:VUFSPQ}(a-l). Here $p_a$, $p_b$, $p_c$, $q_a$, $q_b$ and $q_c$ denote active powers of phases $a, b$ and $c$ and reactive power at phase $a, b$ and $c$, respectively. 
As it can be seen, these sensitivities vary significantly along the nodes and phases and differ for different unbalance metrics. From the plots, it can be observed that the VUFS and PVURS vary over a wide range of values, whereas the LVURS remain fairly within a small range. It suggests that LVUR is equally sensitive to the power injections irrespective of their location in the network, whereas VUF and PVUR are quite sensitive with respect to the locations of the injections. 
\subsection{Sensitivity with change in active and reactive injections}
\label{sec:sense_sense}
\begin{figure}[!htbp]
    \vspace{-1em}
    \centering
    \subfloat[$\partial (\Xi^\text{VUF})/ \partial p$]{ \includegraphics[width=0.4\linewidth]{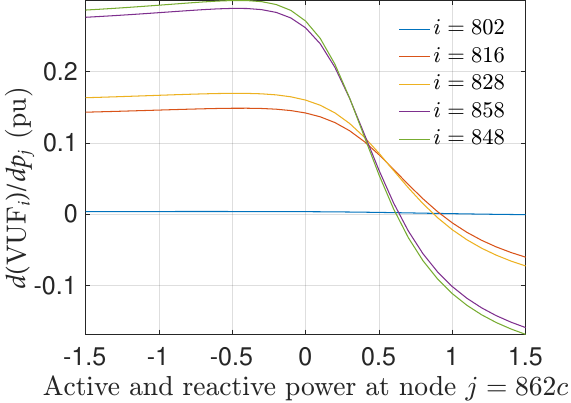}} 
    \subfloat[$\partial (\Xi^\text{VUF})/ \partial q$]{ \includegraphics[width=0.4\linewidth]{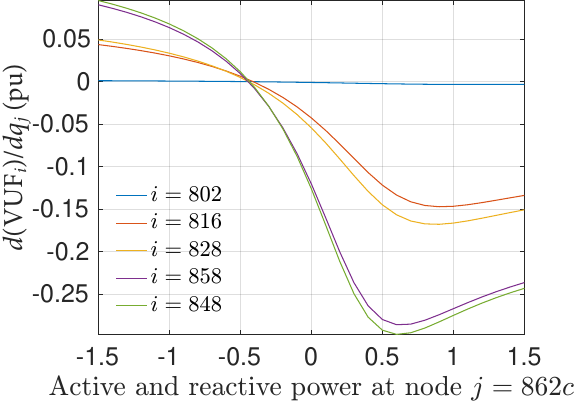}}  
    \caption{VUF sensitivity with varying $p, q$ at node 862, phase c.}
    \label{fig:VUFSwPQ}
        \vspace{-1em}
\end{figure}

\begin{figure}[!htbp]
    \vspace{-1em}
    \centering
    \subfloat[$\partial (\Xi^\text{PVUR})/ \partial p$]{ \includegraphics[width=0.4\linewidth]{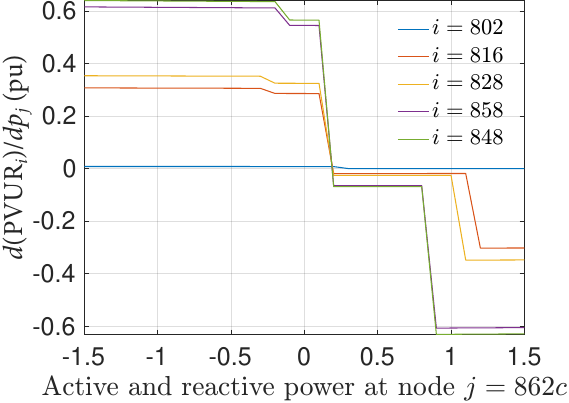}} 
    \subfloat[$\partial (\Xi^\text{PVUR})/ \partial q$]{ \includegraphics[width=0.4\linewidth]{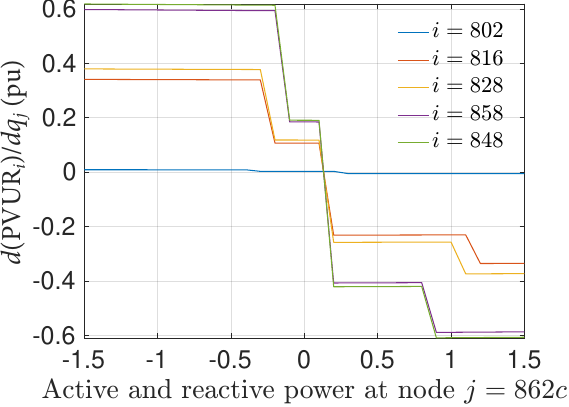}}  
    \caption{PVUR sensitivity with varying $p, q$ at node 862, phase c.}
    \label{fig:PVURswPQ}
        \vspace{-1em}
\end{figure}

\begin{figure}[!htbp]
 \vspace{-1em}
    \centering
    \subfloat[$\partial (\Xi^\text{LVUR})/ \partial p$]{ \includegraphics[width=0.4\linewidth]{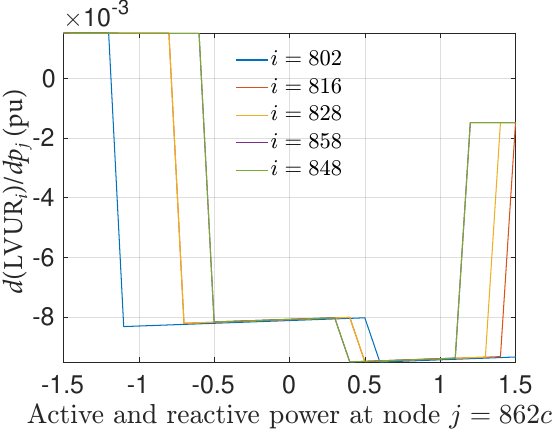}} 
    \subfloat[$\partial (\Xi^\text{LVUR})/ \partial q$]{ \includegraphics[width=0.4\linewidth]{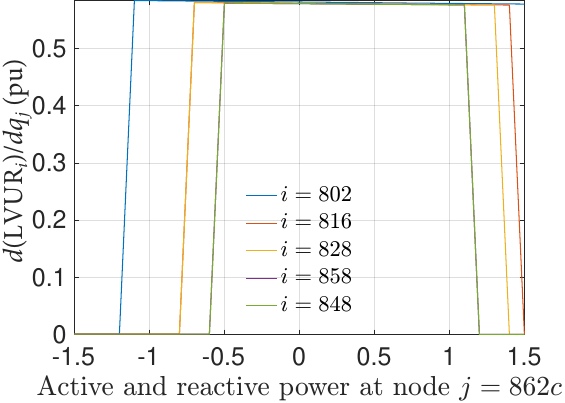}}  
    \caption{LVUR sensitivity with varying $p, q$ at node 862, phase c.}
    \label{fig:LVURSwPQ}
        \vspace{-1em}
\end{figure}

Figures~\ref{fig:VUFSwPQ}, \ref{fig:PVURswPQ} and \ref{fig:LVURSwPQ} report the evolution of the VUF, PVUR, and LVUR sensitivities with change in the injections of one particular node and phase. We present the sensitivities of nodes 802, 816, 828, 848, and 858 with respect to the active and reactive injections at node 862, phase $c$. The active and reactive powers are varied together by a multiplication factor to the nominal injection at that bus from -1.5 to 1.5 with intervals of 0.1. As seen from the plot, the sensitivity of VUF with respect to active and reactive powers is smooth, whereas the ones for PVUR and LVUR are discontinuous. This is because the PVUR and LVUR sensitivities derived in Sec.\ref{eq:PVURS_eq} and \ref{eq:LVURS_eq} were not smooth. The plot also shows that these sensitivities are changing rapidly with the change in the injections, which tells that the linearization of the VUF, PVUR, and LVUR is only valid in the close vicinity of the operating point.  

\subsection{Estimating VUF, PVUR, and LVUR using Sensitivities}
\begin{figure}[!htbp]
    \centering
    \subfloat[Estimated VUF for a perturbed state.]{ \includegraphics[width=0.9\linewidth]{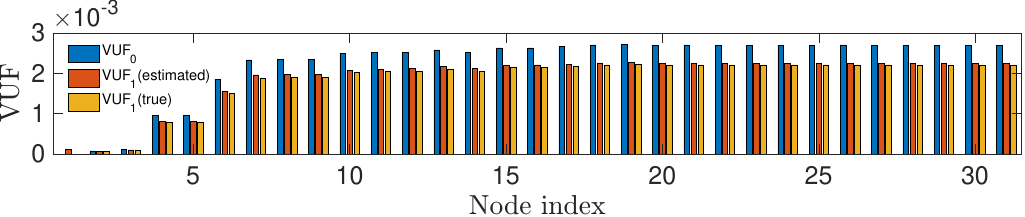}} \\
        \vspace{-0.5em}
    \subfloat[Estimated PVUR for a perturbed state.]{ \includegraphics[width=0.9\linewidth]{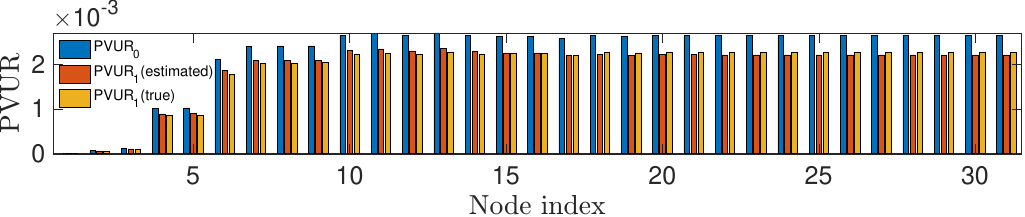}}\\ 
        \vspace{-0.5em}
    \subfloat[Estimated LVUR for a perturbed state.]{ \includegraphics[width=0.9\linewidth]{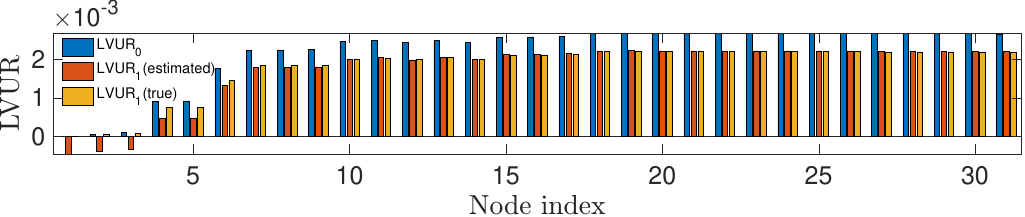}}
    \caption{Underloading case as state 1 (0.8 $\times$ nominal load).}
    \label{fig:underload}
    \vspace{-1em}
\end{figure}

\begin{figure}[!htbp] 
        \vspace{-1em}
    \centering
    \subfloat[VUF estimation for a perturbed state.]{ \includegraphics[width=0.9\linewidth]{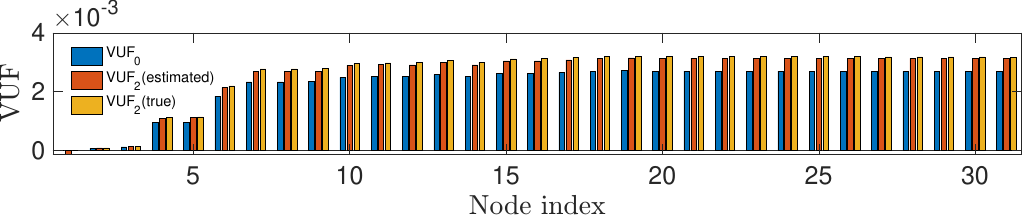}} \\
        \vspace{-0.5em}
    \subfloat[PVUR estimation for a perturbed state.]{ \includegraphics[width=0.9\linewidth]{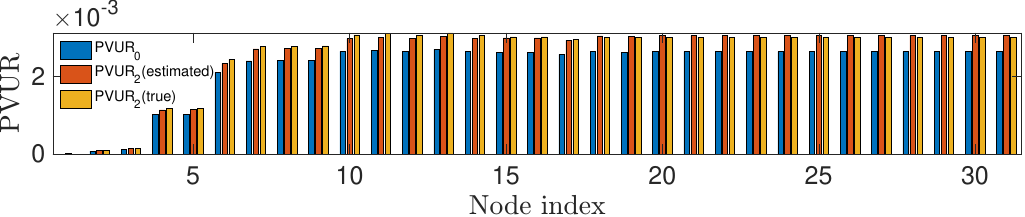}}\\ 
        \vspace{-0.5em}
    \subfloat[LVUR estimation for a perturbed state.]{ \includegraphics[width=0.9\linewidth]{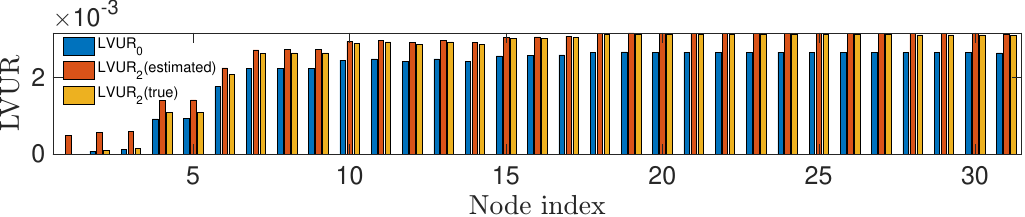}}
    \caption{Overloading case as state 2 (1.2 $\times$ nominal load).}
    \label{fig:overload}
    \vspace{-1em}
\end{figure}

This section shows how the sensitivities are used to estimate the VUF, PVUR, and LVUR for another operating point ${\mathbf{x}}$ using a known operating point $\hat{\mathbf{x}}$ using \eqref{eq:linearization}. For this simulation,  $\hat{\mathbf{x}}$ is used as the nominal loading (State 0) as an operating point. We show estimation for two cases: (i) \textit{underloading}: new injections are 0.8 times nominal load, and (ii) \textit{overloading}: new injections are 1.2 times nominal load. The results for the two cases (state 1 and 2) are compared against the true value (computed with exact power flow) and shown in the barplot in Figs.~\ref{fig:underload} and \ref{fig:overload} for VUF, PVUR and LVUR. As can be seen, the estimated values (using \eqref{eq:linearization}) closely match the true AC power flow values for both cases.

\subsection{Phase balancing application}
\label{sec:phase_bal_results}
\begin{figure}[!htbp]
    \centering
    \includegraphics[width=1\linewidth]{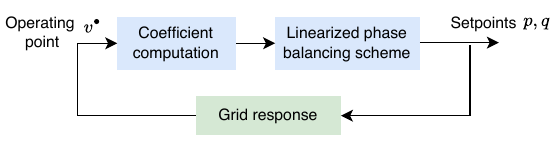}
    \caption{Flow diagram for the control scheme.}
    \label{fig:control_flow}
\end{figure}
This section illustrates how the sensitivity-based linear phase balancing formulation in Sec.~\ref{sec:phase_balancing_application} can improve unbalance metrics by reallocating power injections per phase per node. The objective is to minimize a specific unbalance metric of interest (i.e., VUF, PVUR, or LVUR) with respect to the voltage constraints and the bounds of the unbalance metric. Figure~\ref{fig:control_flow} shows the flow diagram for the phase balancing scheme. As it is shown, using the known operating point (either from the measurements followed or by a state estimation process), first the coefficients for the voltage magnitudes and the unbalance metrics are computed, then they are used in the phase balancing scheme in \eqref{eq:phase_balance_op}, then the setpoints are realized and the response is observed in terms of actualized voltages, which acts as the operating point for the next time step.
\begin{figure}[!htbp]
    \centering
    \subfloat[VUF with and without balancing.]{ \includegraphics[width=0.75\linewidth]{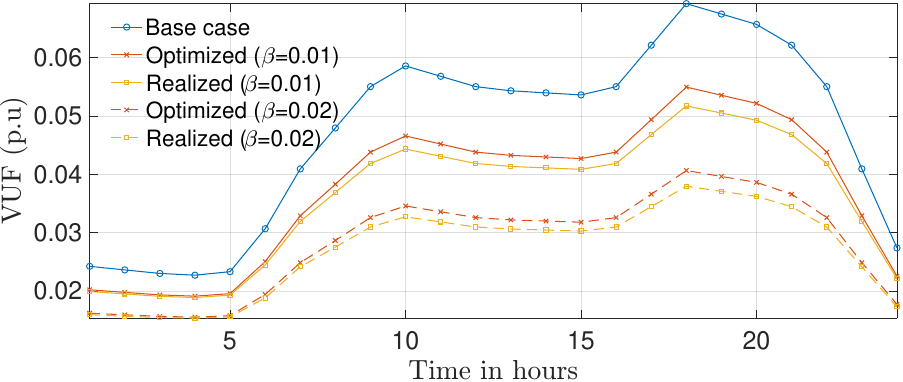} \label{eq:vuf_bal}} \\
    \subfloat[PVUR with and without balancing.]{ \includegraphics[width=0.75\linewidth]{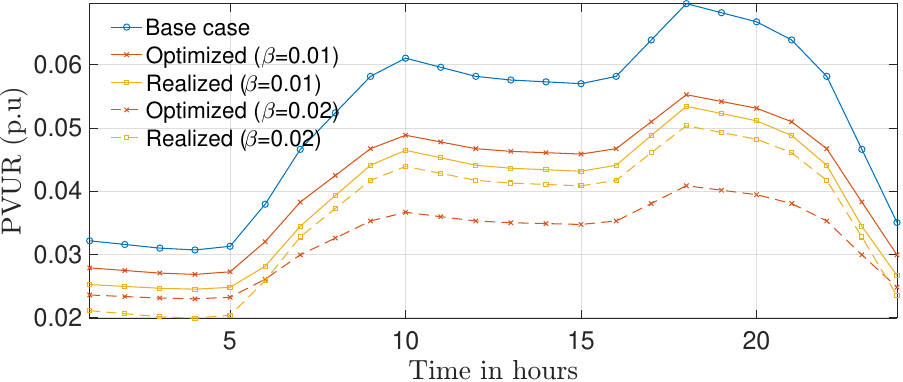}\label{eq:pvur_bal}} \\ 
     \subfloat[LVUR with and without balancing.]{ \includegraphics[width=0.75\linewidth]{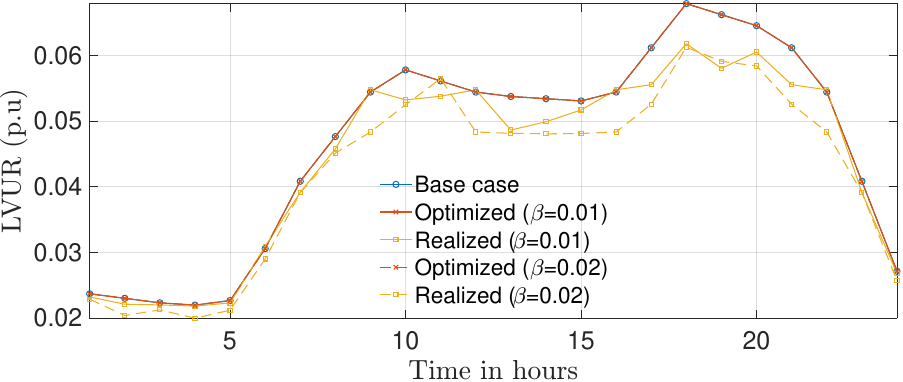}\label{eq:lvur_bal}}
    \caption{Voltage balancing results for $\beta = $ 0.01 and 0.02. \textit{Base case} refers to no phase balancing, \textit{Optimized} refers to the results after solving phase balancing problem in \eqref{eq:phase_balance_op} and \textit{Realized} is the actualized value after solving AC power flow \textit{ex-post} with outputs of phase balancing problem.}
    \label{fig:Voltage_balancing}
\end{figure}

\begin{table}[!htbp]
    \centering
    \renewcommand{\arraystretch}{0.7}
    \caption{Phase balancing results.}
        \resizebox{0.5\textwidth}{!}{%
    \begin{tabular}{|c|c|c|c|}
    \hline
    \bf{Metrics} & \bf Without & \bf Balancing & \bf Balancing \\
     & \bf \bf balancing  & \bf $\beta =  0.01$ &  $\beta =  0.02$\\ 
     \hline
     \textbf{VUF} $(\sum_i\Xi_i^{\text{VUF}})$ & 1.1 & 0.87 (-23.2\%) & 0.65 (-42.47\%)\\
       \hline
     \textbf{PVUR} $(\sum_i\Xi_i^{\text{PVUR}})$ & 1.23 & 0.93 (-23.7\%) & 0.87(-29.4\%)\\
       \hline
     \textbf{LVUR} $(\sum_i\Xi_i^{\text{LVUR}})$ & 1.12 & 1.06 (-4.7\%) & 1.01 (-8.9\%)\\
    \hline
    \end{tabular}}
    \label{tab:VU_optimal}
\end{table}

The results are presented for the IEEE-34 network. We assume that the power at the phases within a node can be changed upto $\beta$ times the nominal value at the phase and the total injection at the node remains unchanged. The results are presented for two different $\beta = 0.01$ and $\beta = 0.02$, referring to 1\% and 2\% of the injections that are flexible. The results are presented for the whole day in Figs.~\ref{fig:Voltage_balancing} 
(a-c). In the figure, the plots in blue color refer the base case i.e., without any optimization, red colored curves show the optimized results computed through the linear model in \eqref{eq:phase_balance_op}, and orange curves are the actual achieved values computed by feeding the power regulation decisions in to the AC power flow \textit{ex-post}. As it can be observed from the plots, the voltage balancing indeed helps in improving the balancing metrics. The orange curve (realized metrics) is always below the base case. Although, there is error between the optimized (computed with the linearization) and the actualized true metric - we always achieve a better unbalance metric. This mismatch between the actual and the optimized is due to the linearization error. The improvement in the unbalance metrics is summarized in Table~\ref{tab:VU_optimal}, where we can observe that all the unbalance metrics achieve substantial improvements with phase balancing. From the plots and the Table, we observe that using LVUR model is not very effective in reducing the unbalances compared to the VUF and LVUR, this might be because of very small sensitivities of LVUR.


\section{Conclusion}
\label{sec:conclusion}
This work proposed a linear phase balancing scheme for distribution systems to tackle the inherent nonlinearity of voltage unbalance metrics and power flow equations. It presented a new set of sensitivity coefficients related to the voltage unbalance definitions from IEEE, IEC, and NEMA, namely Voltage Unbalance Factors (VUF), Phase Voltage Unbalance Rates (PVUR), and Line Voltage Unbalance Rates (LVUR).

The proposed scheme was validated on an IEEE benchmark network, which showed that the sensitivity of VUF is continuous, whereas that of PVUR and LVUR is discontinuous. It was also found that VUF and PVUR values are quite sensitive to the locations and values of the injections, whereas LVUR does not change significantly with the location of the injection in the network. The proposed linearization was then illustrated in a phase balancing application where the injections among phases were varied within a node while keeping the net injection per node constant. The results showed improvement in voltage unbalance metrics up to 40\% with just 2\% variation in the per phase injections.

Future work will explore the use of unbalance sensitivities for other applications such as phase swapping or phase allocation schemes, resulting in mixed integer problems.

\bibliographystyle{IEEEtran}
\bibliography{bibliography.bib}
\end{document}